\documentclass[prl,twocolumn,aps,amsmath,nofootinbib,superscriptaddress]{revtex4}

\usepackage{graphicx}
\usepackage{bm}
\usepackage{epsfig}
\usepackage{color}
\usepackage{xcolor}

  \newcount\hour \newcount\minute
  \hour=\time \divide \hour by 60
  \minute=\time
  \newcommand{\mydate}{\ \today \ - \number\hour :\ifnum \minute<10 0\fi 
\number\minute}

\def\OMIT#1{}

\newcommand{\nn}{\nonumber}

\newcommand{\bea}{\begin{eqnarray}}
\newcommand{\eea}{\end{eqnarray}}

\newcommand{\beq}{\begin{equation}}
\newcommand{\eeq}{\end{equation}}

\begin{document}


\preprint{ \hbox{MIT-CTP 3469} \hbox{CMU-HEP-04-01} \hbox{CALT-68-2475}
 \hbox{hep-ph/0401188}  }

\title{\boldmath
  Two-Loop Matching Onto Dimension Eight Operators in the Higgs-Glue Sector
}

\author{Duff Neill}
\affiliation{Department of Physics, Carnegie Mellon University,
    Pittsburgh, PA 15213  \vspace{0.3cm}}

\begin{abstract}
  
 This letter presents results for the two-loop matching coefficients for
 the dimension eight operators that contribute to Higgs production via gluon fusion.
The coefficients can be used to calculate the first correction to the infinite top mass 
limit to Higgs production with large transverse momentum at two loops. To date such
processes have been studied at two loop order only in the leading term in the top mass expansion.
These corrections become enhanced in processes with large final state invariant mass, typical of
multijet processes. 
\end{abstract}

\maketitle
 
One significant objective of the Large Hadron Collider (LHC) is to produce and study the Higgs particle. To fully understand this
sector of the standard model, it is important to have precise theoretical control over the observables
associated with the Higgs in a hadronic production environment. 
The dominant mechanism for Higgs production is gluon fusion through a top-quark loop (for review see \cite{Djouadi:2005gi}).
Given that the process starts at one loop, it could easily be enhanced by new physics.
Thus an accurate prediction for the cross section and other associated observables is of vital importance, and
much work has been done with this in mind \cite{Harlander:2002wh,Anastasiou:2002yz,Harlander:2009bw,Pak:2009bx}. 
On the other hand, the fact that the process starts at one loop also means that radiative corrections, which
are known to be large, are more difficult to calculate. 
However, the calculations can be greatly simplified by working in an effective
theory where the top quark has been integrated out.  Doing so effectively reduces the order of the calculation 
by one loop at the cost of introducing errors that are suppressed by inverse powers of the top mass. 
To date, the focus has been on the lowest mass dimension six operator generated, 
\beq
\label{dimsix}
L^6_{eff} = C \frac{H F^{a}_{\mu\nu}F^{a \mu\nu}}{v}. 
\eeq
The two loop result \cite{1991NuPhB.359..283D,1991PhLB..264..440D} for the coefficient $C$ is given by
\beq
C= \frac{g^2}{48\pi^2}+\frac{g^4}{4\pi^4}(\frac{5}{192}C_A+\frac{-1}{64}C_F)
\eeq
For $m_{h} < 2 m_{t}$, this leading order (in inverse  powers of $m_t$) contribution does an
excellent job of approximating the inclusive Higgs production rate, with errors on
the order of a few percent for a light Higgs.
On the other hand, observables requiring a large transverse momenta for the Higgs, like production in association with jets 
or its transverse momentum spectrum, will be more susceptible to larger power corrections. Such processes have been calculated to two loops
in the infinite top mass limit, that is, only including the dimension six operator Eq.~(\ref{dimsix})~\cite{deFlorian:1999zd,Glosser:2002gm}.
The full $m_t$ dependence for Higgs plus jet observables is presently only known at one loop
\cite{DelDuca:2001fn,PhysRevD.67.073003,Campbell:2006xx,Keung:2009bs}. 
To extend these observables to two loops is of considerable difficulty, however, 
the calculation is simplified by working in the effective theory. The mass corrections
to these results can be included by first calculating the matching coefficients to the set
of dimension eight operators, and then using those results to calculate the cross section.
The goal of this paper to present the aforementioned Wilson coefficients.
In a forthcoming paper the results for Higgs plus jet cross section will be presented.

Finally, it is interesting to note that these dimension eight operators 
are responsible for the leading order contribution to an observable that can be used as
a test for new physics effects. In particular, it has been shown \cite{ARZ} that the
ratio of the inclusive $\sigma_{inc}$ to cut  $\sigma_{p_t>p_t^0}$ cross sections
\beq
R=\sigma_{inc}/\sigma_{p_t>p_t^0} \approx R_{SM}
\eeq
is approximately model independent if all the new masses are sufficiently large that the effective field
theory is well behaved.  The corrections to this statement arise from the dimension eight operators.
That is
 \beq
 \delta\equiv 1-R/R_{SM} \propto C_8 C_6
 \eeq
where $C_8$ corresponds to some linear combinations of the Wilson coefficients introduced below.
Thus, if $\delta$ is measured and found to be non-zero, then whether
or not one can conclude there must be light new particles in the spectrum can only be 
determined once one determines if the contribution from the dimension eight operators
is sufficiently small. This calculation will be taken up in a future paper.
\section{The Operator Basis}
Below (Eqs.~(\ref{dimeightbasis1}) to (\ref{dimeightbasis4})) is the list of all possible operators with mass dimension eight
coupling gluons to the Higgs, consistent with requirements of Lorentz and color gauge invariance. After using integration by 
parts to remove any derivatives on the Higgs field, the Bianchi identity was used to remove any remaining relations, thus giving a 
linearly independent basis. A minimal basis consists of four operators, and is given by
\begin{eqnarray}
\label{dimeightbasis1}
O_a &=& \frac{H D_{\alpha}F^{a}_{ \mu\nu}D^{\alpha}F^{a \mu\nu}}{m_t^3}\\
\label{dimeightbasis2}
O_b &=&  \frac{H F^{a}_{ \alpha\nu}D^{\nu}D^{\beta}F^{a}_{\beta \alpha}}{m_t^3}\\
\label{dimeightbasis3}
O_c &=& \frac{H D^{\alpha}F^{a}_{ \alpha\nu}D_{\beta}F^{a \beta\nu}}{m_t^3}\\
\label{dimeightbasis4}
O_d &=& \frac{H F^{a\mu}_{\nu}F^{b\nu}_{\sigma}F^{c\sigma}_{\mu}f^{abc}}{m_t^3}
\end{eqnarray}

Three of the four  operators couple to two gluons and the Higgs,
while the fourth is only involved in process involving three or more gluons. Its color factor
only includes the antisymmetric color structure constants. Note that the basis includes two operators $O_b$ and $O_c$ that can be traded for
operators involiving quark bilinears using the equations of motion. However, calculating off-shell
will allow us to utilize the Low Energy Theorem, giving a computationally 
simpler way to calculate the matching coefficients, as discussed below. 
One can then use the equations of motion to simplify the basis after matching. 

\section{Methods}
\label{sec:II}

Canonical matching involves calculating in the full and effective theory and
then taking the difference to find the matching coefficients.  Alternatively, one can 
extract the matching coefficient by asymptotically expanding the integrals around
hard loop momenta which are taken to be of order  $m_t$ \cite{Smirnov:2002pj} and ignoring other regions
which would cancel in the matching. Calculating in this way reduces the amount of work involved.

\begin{figure}[!t]
\centerline{\scalebox{0.75}{\includegraphics{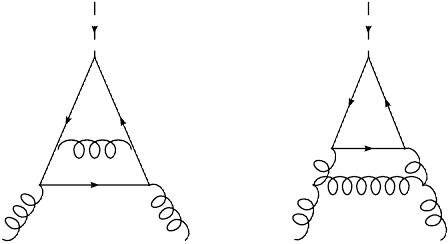}}}
\caption[1]{Representative diagrams for the Higgs-two-gluon Vertex necessary for fixing operators with two gluon Feynman rules.}
\label{Higgs-two-gluon}
\end{figure}

\begin{figure}[!t]
\centerline{\scalebox{0.75}{\includegraphics{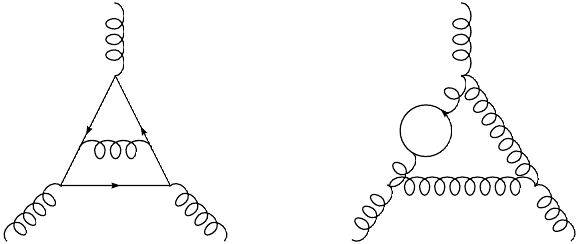}}}
\caption[2]{Representative diagrams for the Three-gluon Vertex necessary for fixing operator the three field strength operator.
Taking a mass derivative with respect to the top quark gives the low energy limit to the Higgs-three-gluon vertex.}
\label{Three-gluon}
\end{figure}

Matching onto the basis requires asymptotically expanding double boxes in the large top mass limit. 
The Low Energy Theorem (LET) for the Higgs (see \cite{Kniehl:1995tn} for an overview)
allows one to reduce the complexity of the calculation. In its basic form, the Low Energy Theorem states that the
amplitude for the process $X \rightarrow Y + H$ can be related to the process $X \rightarrow Y$ as 

\begin{eqnarray*}
lim_{p_h \rightarrow 0} M(X \rightarrow Y + H) =\sum_i \lambda_i m_{q_i} \frac{d}{dm_{q_i}} M(X \rightarrow Y)
\end{eqnarray*}

where $p_h$ is the four momentum of the Higgs boson, and $m_{q_i}$ and $\lambda_i$ are the masses and couplings of the
particles coupling to the Higgs. Diagrammatically this is shown in Figure 3.

\begin{figure}[!t]
\centerline{\scalebox{0.75}{\includegraphics{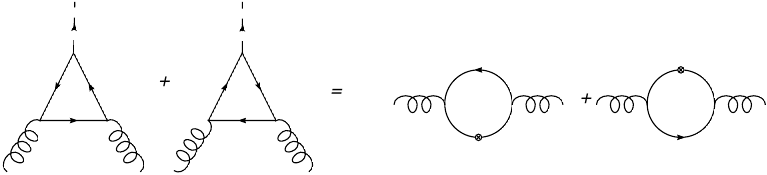}}}
\caption[3]{Illustration  of the low energy theorem. The circles denote mass derivatives and the
higgs momentum vanishes on the left hand side. }
\label{Three-gluon}
\end{figure}

To use the LET in  the matching, one first calculates off shell the corrections to the Higgs--two-gluon 
vertex and match onto three of the operators (This vertex has been investigated in the onshell limit for total Higgs 
production\cite{Dawson:1993qf}), typical diagrams are found in Figure 1. Then one calculates the top-quark contribution to the three-gluon vertex in QCD. Relating this quantity
to the Higgs-three-gluon vertex in the limit of vanishing Higgs four-momentum, one can fix the fourth operator's
matching coefficient. The low energy limit itself is off-shell, hence the inclusion of operators that vanish by the equations of motion. 
The Low Evergy Theorem approach eliminates the need to calculate the 135 two-loop box diagrams 
for the Higgs--three-gluon effective vertex. Instead one need only calculate 57 two-loop triangle diagrams. 

The hard contribution of the integrals is obtained by Taylor expanding in the external momenta. 
The resulting expansion leaves one with a sum of 
bubble diagrams~\cite{Smirnov:2002pj} which are much simpler to evaluate. An efficient method to accomplish the expansion is to first
reduce all integrals to scalar integrals and then performing the Taylor expansion following Tarasov~\cite{Tarasov:1,Tarasov:2}.
In this method, one takes an integral of the form 
\begin{equation}
\label{tensorintegral}
I(s)=\int \prod_{i=1}^{L} d^dk_i  \frac{1}{\prod_{j=1}^{n}(\bar{k}_{j}^2-m_{j}^2)^{\nu_{j}}}e^{i\sum_{l=1}^{L} k_{l}\cdot a_{l}}
\end{equation} 
where $\bar{k}_{j}$ and $m_{j}$ are the momentum and mass associated with the jth propagator. For each loop momentum, 
we have introduced an auxilary vector $a_{l}$ and the exponential factor $e^{i\sum_{l=1}^{L} k_{l}\cdot a_{l}}$. Differentiating 
with respect to the auxiliary vectors allows one to produce any numerator in the loop momenta in the integral from the scalar integral. 
But before differentiating to produce the desired tensor integrals, one passes to the $\alpha$ representation, so the above integral has the form
\bea
I(s,m)=\frac{\Gamma(\sum_{i}^{n}\nu_{i})}{\prod_{i}^{n}\Gamma(\nu_{i})}\int_{0}^{\infty}\prod_{i}^{n}d\alpha_{i} \prod_{i}^{n}\alpha_{i}^{\nu_{i}-1}(D(\alpha))^{-d/2}\nn \\ \!\!\!\times \exp(i\frac{Q(s,\alpha)}{D(\alpha)}-i\sum_{i}^{n}\alpha_{i}m_{i}^2)
\eea
where $s$ are the kinematic invariants formed from the external momenta and auxiliary vectors. $D(\alpha)$ and $Q(s,\alpha)$ are polynomials 
in $\alpha_i$ and $s$, uniquely determined by the topology of the diagram. As noted above, differentiating the integral
with respect to the auxiliary vectors generates the desired numerator. In the $\alpha$-representation the differentiation
generates a polynomial in the external momenta whose coefficients are proportional to scalar integrals
having the same form as $I$, but with shifted spacetime dimension and powers of propagators. This is a simple consequence of the fact that $Q$ is polynomial
in $\alpha_i$ and the kinematic invariants. The shift in spacetime dimension accounts for each derivative bringing down 
an inverse power of $D(\alpha)$. After differentiation, one sets the auxiliary vectors to zero.

Having reduced the diagram to scalar integrals, the Taylor expansion can be performed similarly. The Taylor expansion in the $\alpha$ 
representation is equivalent to writing out the Taylor series in the $\exp(i\frac{Q(s,\alpha)}{D(\alpha)})$ factor, and 
distributing through the $\alpha$ integrations over the terms. Again this results in further shifts in spacetime
dimensions and powers of propagators.

After the reduction to scalar integrals and Taylor expansion, one is left with bubble integrals with propagators
of arbitrary powers, and shifted spacetime dimensions. For the case of the two-loop calculations in the large mass expansion, 
all these integrals were of the form 
\beq
\int \frac{d^{d}k_{1}d^{d}k_{2}}{(k_{1}^{2}-m^{2})^{\nu_1}((k_{1}+k_{2})^2)^{\nu_2}(k_{2}^{2}-m^{2})^{\nu_3}}
\eeq
where a simple analytic result is known for all $\nu_{i}$ and spacetime dimension $d$. 
To insure gauge invariance of the final results, all effective action vertices were computed in the background field gauge, and renormalized
with the $\overline{\text{MS}}$ scheme at the scale $2 m_t$. 

All diagrams were generated with FeynArts\cite{Hahn:2000kx} and then analysized within Mathematica.

\section{Calculational Checks}
\label{sec:IV}

We have checked that we reproduce the known matching results for the dimension six operator to two loops. 
This check works independently of the choice of external states, so we reproduce the matching to the 
dimension six operator in both processes computed. 

From the calculation we can extract the anomolous dimensions of the operator basis and compare it
to known results, thus providing another non-trivial check on the calculation. 
In the method of regions, each region develops infrared and ultraviolet divergences, but only the sum over the
regions contains the divergences (both UV and IR) of the full theory\cite{Smirnov:2002pj}. Thus UV divergences of the
soft regions must cancel with IR divergences of the hard region. With knowledge of the UV divergences
of the effective theory (contained in the anomolous dimensions of the effective operators), and the
one loop matching, one can predict the IR divergences of the hard region. The anomolous dimensions
of the effective operators have been computed before \cite{Gracey:2002rf,Narison1983217} (the operators considered
there were pure QCD operators, but have the same QCD renormalization properties since the higgs field is a color singlet).
Thus it becomes a simple matter to check that the coefficients of the logarithms in the asymptotic expansion
are the one loop matching coefficients times the renormalization factor needed to substract the UV divergences
of the effective theory. Thus schematically if we have in a calculation for the hard region ($\textit{HR}$)
\begin{eqnarray*}
\label{hardregion}
\textit{HR} &=& A_{UV}*(\epsilon_{UV}^{-1}+ Log(\frac{\Lambda^2}{\mu^2}))+ \\ && B_{IR}*(\epsilon_{IR}^{-1}+ Log(\frac{\Lambda^2}{\mu^2})) + Finite,
\end{eqnarray*}
the effective field theory ($\textit{EFT}$) then has  
\begin{eqnarray*}
\label{softregion}
\textit{EFT} &=& C_{UV}*(\epsilon_{UV}^{-1}+ Log(\frac{p^2}{\mu^2}))+ \\ && D_{IR}*(\epsilon_{IR}^{-1}+ Log(\frac{p^2}{\mu^2})) + Finite.
\end{eqnarray*}
Where $\Lambda$ is the hard scale, and $p$ is the effective theory scale. The two are reproducing the full theory
when $B_{IR}=-C_{UV}$. 

The only UV logarithms of the hard region correspond to the top quark mass renormalization. 
The two types of logarithms are easily distinguished by the associated group theory factors due to the 
differing representations of quarks and gluons.

\section{Results}
\label{sec:V}

The effective lagrangian resulting from integrating out the top to this mass order is:

\begin{eqnarray*}
L_{eff}&=&C_{1} \frac{H F^{a}_{\mu\nu}F^{a \mu\nu}}{m_t}+C_{2}\frac{H D_{\alpha}F^{a}_{ \mu\nu}D_{\alpha}F^{a \mu\nu}}{m_t^3}+\\
& & C_{3}\frac{HF^{a\mu}_{\nu}F^{b\nu}_{\sigma}F^{c\sigma}_{\mu}f^{abc}}{m_t^3}+\\
 & & + C_{4}\frac{H D^{\alpha}F^{a}_{ \alpha\nu}D_{\beta}F^{a \beta\nu}}{m_t^3}+C_{5}\frac{H F^{a}_{ \alpha\nu}D^{\nu}D^{\beta}F^{a}_{\beta\alpha}}{m_t^3}\\
\end{eqnarray*}

Where
\begin{eqnarray*}
C_{1}&=&\frac{g^2 \lambda}{48\pi^2}+\frac{g^4 \lambda}{4\pi^4}(\frac{5}{192}C_A-\frac{1}{64}C_F)\\
C_{2}&=&\frac{-7 g^2 \lambda}{2880\pi^2}+\frac{-g^4 \lambda}{4\pi^4}(\frac{29}{34560}C_A\\ &&+\frac{19}{8640}C_F+\frac{-7}{1920}(C_F)Log(\frac{\pi e^{\gamma} m_t^2}{\mu^2}))\\
C_{3}&=&-\frac{g^3 \lambda}{240\pi^2}+\frac{g^5\lambda}{6\pi^4}(\frac{1}{14400}C_A-\frac{13}{1920}C_F\\ &&-\frac{1}{320}(C_A+3C_F)Log(\frac{\pi e^{\gamma} m_t^2}{\mu^2}))\\
C_{4}&=&\frac{g^2\lambda}{1440\pi^2}+\frac{g^4 \lambda}{2\pi^4}(\frac{-101}{691200}C_A+\frac{1}{3240}C_F\\ &&+\frac{-1}{17280}(29C_A-9C_F)Log(\frac{\pi e^{\gamma} m_t^2}{\mu^2}))\\
C_{5}&=&\frac{g^2\lambda}{80\pi^2}+\frac{g^4 \lambda}{\pi^4}(\frac{1169}{518400}C_A+\frac{73}{51840}C_F\\ &&+\frac{-1}{17280}(56C_A-81C_F)Log(\frac{\pi e^{\gamma} m_t^2}{\mu^2}))\\
\end{eqnarray*}
$\lambda=\frac{m_t}{v}$ is the yukawa coupling to the top quark. One takes $m_t$ in what ever scheme one renormalizes the
hard region. Then one runs the operators to the low scale using a scheme that is consistent with the scheme used in the 
matching. For simplicity the coefficients are listed in the $MS$ scheme.

\section{Conclusion}
\label{sec:VI}

We have presented the order $\alpha_s^2(m_h/m_t)^3$ Lagrangian coupling the Higgs directly to
gluons. This basis will prove useful in understanding the higher order
QCD corrections to Higgs production in association with jets, where the range of validity
of the standard effective field theory begins to break down due to large final state invariant masses. 
In a forth coming paper the basis will be used to examine the gluon induced Higgs production with a large
transverse momentum observable at the LHC. 

\section{Acknowledgements}
The author would like to thank Ira Rothstein for useful discussions and suggesting this project. 
Diagrams were made with Jaxodraw \cite{Binosi:2003yf}. Work supported by DOE contracts DOE-ER-40682-143 and DEAC02-6CH03000 and partial support from NSF grant PHY-0705682.

\section{Note Added:}
The appearance of \cite{Harlander:2013oja} lead the author to revisit his calculation of the matching coefficients, since there was a discrepancy in the leading order matching of the $HFFF$ operator between a previous version of this paper and reference \cite{Harlander:2013oja}. This discrepancy could be traced to a inconsistent sign convention in the covariant derivative and the field strength tensor used to derive the feynman rules in the effective theory in the previous version. Fixing this inconsistency brought the results into line with \cite{Harlander:2013oja} at one loop, and also changed the two loop matching. The full theory calculation remained unaffected. In the previous version of the paper, the erronous value of $C_3$ was given as:
\begin{eqnarray*}
C_{3}&=&\frac{g^3 \lambda}{180\pi^2}+\frac{g^5\lambda}{6\pi^4}(\frac{49}{9600}C_A+\frac{37}{5760}C_F\\ &&+\frac{-1}{320}(C_A-4C_F)Log(\frac{\pi e^{\gamma} m_t^2}{\mu^2}))
\end{eqnarray*}

\vspace{-.4cm}

\bibliographystyle{h-physrev}
\bibliography{dim_eight_matching_paper}

\begin{thebibliography}{10}

\bibitem{Djouadi:2005gi}
A.~Djouadi,
\newblock Phys. Rept. {\bf 457}, 1 (2008), hep-ph/0503172.

\bibitem{Harlander:2002wh}
R.~V. Harlander and W.~B. Kilgore,
\newblock Phys. Rev. Lett. {\bf 88}, 201801 (2002), hep-ph/0201206.

\bibitem{Anastasiou:2002yz}
C.~Anastasiou and K.~Melnikov,
\newblock Nucl. Phys. {\bf B646}, 220 (2002), hep-ph/0207004.

\bibitem{Harlander:2009bw}
R.~V. Harlander and K.~J. Ozeren,
\newblock (2009), arXiv:0907.2997.

\bibitem{Pak:2009bx}
A.~Pak, M.~Rogal, and M.~Steinhauser,
\newblock (2009), arXiv:0907.2998.

\bibitem{1991NuPhB.359..283D}
S.~{Dawson},
\newblock Nuclear Physics B {\bf 359}, 283 (1991).

\bibitem{1991PhLB..264..440D}
A.~{Djouadi}, M.~{Spira}, and P.~M. {Zerwas},
\newblock Physics Letters B {\bf 264}, 440 (1991).

\bibitem{deFlorian:1999zd}
D.~de~Florian, M.~Grazzini, and Z.~Kunszt,
\newblock Phys. Rev. Lett. {\bf 82}, 5209 (1999), hep-ph/9902483.

\bibitem{Glosser:2002gm}
C.~J. Glosser and C.~R. Schmidt,
\newblock JHEP {\bf 12}, 016 (2002), hep-ph/0209248.

\bibitem{DelDuca:2001fn}
V.~Del~Duca, W.~Kilgore, C.~Oleari, C.~Schmidt, and D.~Zeppenfeld,
\newblock Nucl. Phys. {\bf B616}, 367 (2001), hep-ph/0108030.

\bibitem{PhysRevD.67.073003}
V.~Del~Duca, W.~Kilgore, C.~Oleari, C.~Schmidt, and D.~Zeppenfeld,
\newblock Phys. Rev. D {\bf 67}, 073003 (2003).

\bibitem{Campbell:2006xx}
J.~M. Campbell, R.~K. Ellis, and G.~Zanderighi,
\newblock JHEP {\bf 10}, 028 (2006), hep-ph/0608194.

\bibitem{Keung:2009bs}
W.-Y. Keung and F.~Petriello,
\newblock (2009), arXiv:0905.2775.

\bibitem{ARZ}
C.~Arnesen, I.~Z. Rothstein, and J.~Zupan,
\newblock (2008), arXiv:0809.1429.

\bibitem{Smirnov:2002pj}
V.~A. Smirnov,
\newblock Springer Tracts Mod. Phys. {\bf 177}, 1 (2002).

\bibitem{Kniehl:1995tn}
B.~A. Kniehl and M.~Spira,
\newblock Z. Phys. {\bf C69}, 77 (1995), hep-ph/9505225.

\bibitem{Dawson:1993qf}
S.~Dawson and R.~Kauffman,
\newblock Phys. Rev. {\bf D49}, 2298 (1994), hep-ph/9310281.

\bibitem{Tarasov:1}
O.~V. Tarasov,
\newblock Nucl. Phys. {\bf B480}, 397 (1996), hep-ph/9606238.

\bibitem{Tarasov:2}
O.~V. Tarasov,
\newblock Phys. Rev. {\bf D54}, 6479 (1996), hep-th/9606018.

\bibitem{Hahn:2000kx}
T.~Hahn,
\newblock Comput. Phys. Commun. {\bf 140}, 418 (2001), hep-ph/0012260.

\bibitem{Gracey:2002rf}
J.~A. Gracey,
\newblock Nucl. Phys. {\bf B634}, 192 (2002), hep-ph/0204266.

\bibitem{Narison1983217}
S.~Narison and R.~Tarrach,
\newblock Physics Letters B {\bf 125}, 217  (1983).

\bibitem{Binosi:2003yf}
D.~Binosi and L.~Theussl,
\newblock Comput. Phys. Commun. {\bf 161}, 76 (2004), hep-ph/0309015.

\bibitem{Harlander:2013oja}
R.~V. Harlander and T.~Neumann,
\newblock Phys.Rev. {\bf D88}, 074015 (2013), 1308.2225.

\end{thebibliography}

\newpage

\end{document}